\documentclass{aastex631}
\usepackage{xcolor}
\usepackage{multirow}
\usepackage{amsmath}
\usepackage[linesnumbered,ruled,vlined]{algorithm2e}

\SetCommentSty{mycommfont}

\begin{document}
\title{BATSRUS GPU: Faster-than-Real-Time Magnetospheric Simulations with a Block-Adaptive Grid Code}
\author{Yifu An}
\affiliation{University of Michigan}
\author{Yuxi Chen}
\affiliation{University of Michigan}
\author{Hongyang Zhou}
\affiliation{Boston University}
\author{Alexander Gaenko}
\affiliation{University of Michigan}
\author{G\'abor T\'oth}
\affiliation{University of Michigan}



\begin{abstract}
BATSRUS, our state-of-the-art extended magnetohydrodynamic code, is the most used and one of the most resource-consuming models in the Space Weather Modeling Framework. It has always been our objective to improve its efficiency and speed with emerging techniques, such as GPU acceleration. To utilize the GPU nodes on modern supercomputers, we port BATSRUS to GPUs with the OpenACC API. Porting the code to a single GPU requires rewriting and optimizing the most used functionalities of the original code into a new solver, which accounts for around 1\% of the entire program in length. To port it to multiple GPUs, we implement a new message passing algorithm to support its unique block-adaptive grid feature. We conduct weak scaling tests on as many as 256 GPUs and find good performance. The program has 50-60\% parallel efficiency  on up to 256 GPUs, and up to 95\% efficiency within a single node (4 GPUs).
Running large problems on more than one node has reduced efficiency due to hardware bottlenecks.
We also demonstrate our ability to run representative magnetospheric simulations on GPUs. 
The performance for a single A100 GPU is about the same as 270 AMD ``Rome'' CPU cores, and it runs 3.6 times faster than real time. 
The simulation can run 6.9 times faster than real time on four A100 GPUs.
\end{abstract}




\section{Introduction}
The Space Weather Modeling Framework (SWMF) has been developed for more than 20 years \citep{Toth:2005swmf,Gombosi:2021}. By coupling many physical models and computational tools, it enables high-performance yet flexible simulation of the physics from the upper solar chromosphere to Earth's upper atmosphere or to the outer heliosphere. The Block-Adaptive-Tree Solar-wind Roe-type Upwind Scheme (BATSRUS) code models some of the most resource-consuming domains in the SWMF. It is a multi-physics solver with support for various forms of the magnetohydrodynamic (MHD) equations and deals with disparate spatial scales using a block-adaptive grid
\citep{Toth:2012swmf}.
It has always been an important goal to make BATSRUS computationally efficient. A core part of BATSRUS was rewritten into the Block-Adaptive-Tree Library (BATL), a general and efficient toolkit for Adaptive Mesh Refinement (AMR), message passing (communication among neighboring grid blocks), and load balancing \citep{Toth:2012swmf}. BATL was originally designed to run in parallel on multiple CPU cores with the Message Passing Interface (MPI). This model of parallelism has excellent scalability but limits the maximum problem size on systems with limited memory, due to the storage of duplicated arrays on each MPI process. 
To address this issue, \citet{Zhou:2020} implemented a hybrid parallelism with OpenMP and MPI. The number of cores on which BATSRUS has good weak scalability increases from 16,000 cores (MPI only) to more than 500,000 cores (MPI and OpenMP) on a system with 2GB/core memory.

For over a decade, new supercomputers have started to incorporate computing nodes with Graphics Processing Units (GPUs) that provide a huge number of parallel cores to accelerate computationally intensive applications.
The advancements in hardware call for new strategies of parallelism. For the sheer size of BATSRUS, which has more than 200,000 lines of modern Fortran, it is not practical to port it to GPUs with CUDA. Our choice of approach for GPU porting is the Fortran OpenACC API, which provides an excellent balance between good performance and minimal code changes.

OpenACC has proved to be efficient and successful when it comes to porting large Fortran codes to GPU. \citet{Caplan:2019} ported their well-established solar MHD code, Magnetohydrodynamic Algorithm outside a Sphere (MAS), with OpenACC, which has around 50,000 lines of Fortran. 
They added OpenACC directives as long as only 1.5 \% of the original code and modified less than 5\% of it, mainly due to a lack of features in OpenACC (at the time of their work) and miscellaneous optimizations. They found that their OpenACC-accelerated code had excellent compact performance, as 1 node with 4$\times$V100 GPUs ran faster than 8 nodes with 40$\times$Skylake CPU cores. \citet{Zanna:2024} ported to the GPU a state-of-the-art Fortran code for relativistic MHD, Eulerian Conservative High Order (ECHO), also with OpenACC.
They reported a nearly 16-fold speedup on a node with 4 V100 GPUs compared to one with 32 IBM POWER9 cores. They conducted weak scaling tests on up to 256 nodes of 4$\times$V100 GPUs and found a drop in parallel efficiency of up to 30\% as the number of GPUs increased.
The recent addition of compiler support for standard language parallelism in accelerated computing, such as {\tt DO CONCURRENT} (DC) loops in Fortran, provides an alternative to OpenACC.
\citet{Caplan:2023} refactored OpenACC-accelerated MAS into a Fortran-only code accelerated by DC loops, and later used OpenACC directives for memory management.
They found that the Fortran-only code had worse performance on GPUs compared to the original OpenACC-accelerated code, mainly due to the compiler not being able to manage memory efficiently.
The modified code using DC for acceleration and OpenACC directives for memory management had performance similar to the OpenACC-accelerated code, but excelled in portability.

While it is already possible to forecast space weather based on BATSRUS simulations, the quality of the forecast depends on how finely the simulation domain is resolved, what physical models are incorporated, etc.
Increasing the quality of the forecast often means increasing the computational cost of the simulation.
In particular, running an ensemble of simulations can provide improved probabilistic forecasts. This requires a lot of computational resources, so we have to accelerate the simulation as much as possible on the current computational platformss.
The unique block-adaptive-grid feature of BATSRUS creates particular challenges for efficient communication among multiple GPUs, which is addressed in this paper.
This paper discusses our approach to port BATSRUS to one GPU first and then to multiple GPUs communicating via MPI and reports the performance of the resulting accelerated code.

The rest of this paper is organized as follows.
Section \ref{sec:BATL} describes the block-adaptive grid of BATSRUS so that the reader can appreciate the unique challenges of porting BATSRUS to the GPU. 
Section \ref{sec:Porting} details the general strategies to implement the single-GPU version of BATSRUS. Section \ref{sec:Algorithm} introduces the message passing algorithm among multiple CPU cores that prevents parallelization and the new highly parallel algorithm developed for communication among multiple GPUs. The performance of the code on single and multiple GPUs is demonstrated by weak and strong scaling tests in section \ref{sec:tests}. Finally, we conclude with section \ref{sec:Conclusions}.


\section{Block-Adaptive Grid Code on CPUs}
\label{sec:BATL}

The disparate scales in space physics simulations require MHD solvers to emphasize some regions while overlooking the others. Many MHD solvers \citep{Linker:2003,Odstrcil:2003enlil,lyon:2004} use a non-uniform but static grid, some \citep{Janhunen:1996} use a cell-based adaptive approach, while BATSRUS employs a block-adaptive grid (\citet{vanderHolst:2008} adopts a similar approach). 
We refer readers to \citet{Toth:2012swmf} for a discussion of the advantages of using a block-adaptive grid approach. 
In BATSRUS, this strategy is implemented by BATL. BATL enables the user to create Cartesian, rotated Cartesian, cylindrical and spherical adaptive grids in 1, 2, or 3 spatial dimensions. Non-Cartesian grids (cylindrical, spherical, etc.) use generalized  coordinates. 
In generalized coordinates, all grids have the shape of a rectangular prism, consisting of grid blocks. All grid blocks are similar, i.e. each grid block has the same number of grid cells in each dimension, although the sizes of the grid blocks in the domain of simulation may vary. 
A grid block can be refined or coarsened depending on the adaptive mesh refinement (AMR) criteria of the user's choice. The ratio of refinement can only be 2, but the resolution change can take place in 1, 2, or 3 dimensions, so a refined block can have 2, 4, or 8 siblings, while 2, 4, or 8 blocks are coarsened into 1. 
After an AMR, a load balancing module maintains a similar number of blocks on each MPI process by redistributing the blocks.
BATL maintains a tree to keep track of the parents/siblings, as well as neighbors of the blocks.
This tree resides on all MPI processes and is updated everywhere as soon as the grid or local blocks change, for example after an AMR or load balancing.
A message passing module, also using MPI, handles the communication among blocks that reside on different processes. For two physically adjacent blocks that reside on different processes, a layer of ghost cells surrounding the receiving block receives information from the neighboring block. For every pair of processes, message exchange among multiple blocks can take place. Messages can have various sizes if the grid is refined locally. For the sake of efficiency, BATL implements single-buffer communication for every pair of processes. The block index and the index range of the receiving cells are sent together with ghost cell values, and messages are packed one after another. 
Note that the block-adaptive grid feature of BATSRUS poses unique challenges when we port BATSRUS to GPUs.
Section \ref{sec:Porting} provides a detailed discussion.

Algorithm~\ref{solver_cpu} shows the basic structure of the BATSRUS solver.
Here $n_V$ is the number of state variables in each cell, $n_I$, $n_J$, $n_K$ are the numbers of cells in 3 dimensions, $n_B$ is the number of blocks, and $n_S$ the number of time steps.
In the BATSRUS source code, all arrays have a name that indicates the indexes.
For example, State\_VGB contains the current state of the solution indexed by variables (V), grid cells including ghost cells (G) and blocks (B). We assume 2 ghost cells are used in each direction, so the cell index in the $x$ direction is from $-1$ to $n_I+2$, et cetera. The letter C means grid cell centers without ghost cells, X, Y, and Z refer to the faces, P means indexed by processes, and I is a generic index.

First, BATSRUS allocates the arrays State\_VGB (MHD state over the grid), LeftState\_VX \ldots  RightState\_VZ (state extrapolated to the X face from the left $\ldots$ Z face from the right), Flux\_VX $\ldots$ Flux\_VZ (fluxes through the X$\ldots$ Z faces) and Source\_VC (the source terms at the cell centers). Note that only State\_VGB stores information for all the blocks; the other arrays are reused over and over for each block. In the simplified time loop, the solution is advanced for each block in four stages: (1) the left and right face values are interpolated from cell center values, (2) the fluxes are calculated from face values, (3) source terms are obtained, and (4) the solution is updated. 
After finishing the loop over the blocks, the ghost cells are filled in from neighboring blocks and from the boundary conditions. Then the next time step starts.

\section{Porting BATSRUS to One GPU}
\label{sec:Porting}

\begin{algorithm}[t]
\label{solver_cpu}
\caption{BATSRUS solver on CPU.}
\tcc{Allocate variables}
allocate(State\_VGB($n_V,-1:n_I+2,-1:n_J+2,-1:n_K+2,n_B)$)\;
allocate(LeftState\_VX$(n_V,n_I+1,n_J,n_K),\ldots$
RightState\_VZ$(n_V,n_I,n_J,n_K+1)$)\;
allocate(Flux\_VX$(n_V,n_I+1,n_J,n_K),\ldots$
Flux\_VZ$(n_V,n_I,n_J,n_K+1)$)\;
allocate(Source\_VC$(n_V,n_I,n_J,n_K))$\;
\tcc{Perform $n_S$ time steps}
\For{$i_S=1,n_S$}{
\tcc{Loop over $n_B$ grid blocks}
\For{$i_B=1,n_B$}{
   \tcc{From State\_VGB(:,:,:,:,$i_B$) calculate LeftState\_VX $\ldots$}
   call calc\_face\_value($i_B$)\;
   \tcc{From LeftState\_VX $\ldots$ calculate Flux\_VX $\ldots$}
   call calc\_face\_flux($i_B$)\;
   \tcc{From State\_VGB(:,:,:,:,$i_B$) calculate Source\_VC}
   call calc\_source($i_B$)\;
   \tcc{From Source\_VC, Flux\_VX $\ldots$ 
       update State\_VGB($:,1:n_I,1:n_J,1:n_K,i_B$)}
   call update\_state($i_B$)\;
} 
\tcc{Fill ghost cells of State\_VGB}
call exchange\_messages\;
}
\tcc{Write restart files to disk}
call write\_restart\_files\;

\end{algorithm}

\begin{algorithm}[t]
\label{solver_gpu}
\caption{BATSRUS solver on GPU.}
\tcc{Allocate variables on CPU and GPU}
\textcolor{red}{!\$acc declare create(State\_VGB, Flux\_VXB, Flux\_VYB, Flux\_VZB)}\\
allocate(State\_VGB($n_V,-1:n_I+2,-1:n_J+2,-1:n_K+2,n_B)$)\;
allocate(Flux\_VXB$(n_V,n_I+1,n_J,n_K,n_B), \ldots)$)\;
\tcc{Perform $n_S$ time steps}
\For{$i_S=1,n_S$}{
\tcc{Loop over $n_B$ grid blocks in parallel}
\textcolor{red}{!\$acc loop gang independent}\\
\For{$i_B=1,n_B$}{
\tcc{Loop over X grid cell faces in parallel}
\textcolor{red}{!\$acc loop vector collapse(3) independent}\\
\For{$i=1,n_I+1;j=1,n_J;k=1,n_K$}{
  \tcc{Calculate X face values and Flux\_VXB($:,i,j,k,i_B$)}
  call get\_flux\_x($i,j,k,i_B$)\;
}
\tcc{Loop over Y and Z grid cell faces in parallel}
$\ldots$\;
\tcc{Loop over grid cell centers in parallel}
\textcolor{red}{!\$acc loop vector collapse(3) independent}\\
\For{$i=1,n_I; j=1,n_J; k=1,n_K$}{
\tcc{Calculate source and update State\_VGB(:,i,j,k,$i_B$)}
call update\_cell($i, j, k, i_B$)\;
}
}
\tcc{Fill ghost cells of State\_VGB}
call exchange\_messages \\
}
\tcc{Copy State\_VGB to CPU and write restart files to disk}
\textcolor{red}{!\$acc update host (State\_VGB)}\\
call write\_restart\_files\\
\end{algorithm}

OpenACC directives can be added to the original source code to make it run on GPUs. The most important OpenACC directives are the following: declaring the variables that should exist in the GPU memory, moving variables from the CPU to the GPU or vice versa, and making loops parallel on the GPU. 

\subsection{Memory management}

While exploring OpenACC capabilities, we soon realized that moving variables, even just a few bytes, between the GPU and CPU is very slow, so we need to be able to advance time steps purely on the GPU. This means that all variables related to updating the solution over multiple time steps must be declared and updated on the GPU. Since disk access is only possible from the CPU, we need to move data between the GPU and CPU for reading input files, saving plot files and restart files. These operations should be done relatively infrequently (typically less than once per 100 time steps) to minimize the impact on performance.
We also opted, for now, to perform the adaptive mesh refinements on the CPU. This reduces the amount of code changes needed and the storage requirements on the GPU. Doing AMR on the CPU has little impact on performance as long as it is done infrequently, which is typical for most of our applications.

\subsection{Loop ordering}

The amount of parallelism due to the number of grid cells (around 500 to 1000) in a single grid block is not sufficient to keep the GPU busy. On the other hand, because all grid blocks have the same number of grid cells, increasing the number of cells per block sacrifices the flexibility to overlook less interesting regions in the domain. This means that we need to write parallel loops over the grid cells of multiple grid blocks. 
We use a two-level parallelization strategy, i.e. the \emph{gang parallel} loop to parallelize grid blocks, and the \emph{vector parallel} loop to parallelize grid cells. We note that the operations applied to different grid blocks can be different, while the operations on the grid cells of the same grid block are usually the same.



The amount of cache memory is rather limited on the V100 GPUs (a V100 has 6144 KB of L2 cache). This means that we need to use scalars and small arrays inside the loops. 
On the other hand, the only way to pass information between two gang loops is to save the information for all blocks.
This is different from the CPU code, where we have many arrays with cell indexes but no block index, and these transfer information among various operations (face value calculation, flux calculation, source term calculation, cell update) that are performed one by one for a single grid block.

The outlines of the new solver of BATSRUS are shown in Algorithm~\ref{solver_gpu}. 
BATSRUS allocates State\_VGB, Flux\_VXB, Flux\_VYB, and Flux\_VZB on both the CPU and GPU using the
\emph{\$acc declare create} directive.
Note that the Flux arrays now have 
a block index; on the other hand,  
the LeftState, RightState, and Source arrays have been eliminated. The face states are calculated and used immediately to obtain the fluxes at the faces, while the source terms are calculated and used for the update for each cell center. The solution is moved from the GPU to the CPU using the \emph{\$acc update host} directive only when needed, for example, for writing plots and restart files. 

As we implement more and more features, the content of the global loop becomes increasingly complex. To help the nvfortran compiler produce efficient code, we have devised the following strategy. 
Before compilation, a Perl script checks the input file PARAM.in and identifies logical switches and other simple parameters that remain constant throughout the run.
The script then writes those values into a Fortran module file as Fortran {\tt parameter}s, so that the compiler will skip the unused sections of the code.
For example, if the simulation uses a second-order scheme, the user defines nOrder=2 in PARAM.in. 
The Perl script then reads PARAM.in and declares {\tt integer, parameter::nOrder=2} in the source code before compilation, so that the code segments written for the first-order scheme are not compiled. 
Another astrophysics MHD code that adopts this approach is Athena \citep{Stone:2008}.
While our approach is similar to the use of compiler directives, there is much less change compared to the original source code, and the code readability is also improved. 

\subsection{Implementation}

While the original code is close to being optimal for CPUs, it was not designed for GPUs and thus cannot run efficiently on them. Therefore, we have rewritten the main extended MHD solver completely.
While this may seem like giving up on the minimal code change principle, in fact, the new solver is only around 2,500 lines at this point, which is around 1\% of the total code base of BATSRUS. One reason for the small number of lines is that we only implemented the most used solver options, and another is that we could call/reuse some subroutines that only needed minimal changes from the rest of the code. 
We added only around 800 lines of OpenACC directives and a few hundred lines of new or modified source code in addition to the main GPU solver module.

Having a working CPU implementation makes the GPU porting much easier. The first step after implementing 
a new piece of code intended for the GPU is to test it on a CPU with the compiler's debugging options switched on. Next, the OpenACC declarations of global and local variables are added until the code successfully compiles for the GPU. Next, the execution on the GPU is tested against the original CPU code. If the results are significantly different, detailed debugging is done on the newly implemented source code. Writing out intermediate results from the CPU and GPU code makes this relatively straightforward. This usually reveals some variables that should be declared private for the loop. The next crucial step is checking the performance on the GPU. We often find that the performance gets drastically worse after adding new code. A typical reason is that some variables are not declared on the GPU, and OpenACC attempts to copy the data from the CPU automatically. While this approach works in terms of obtaining correct results, the performance suffers. Once the code produces the same result on the GPU as on the CPU with good performance, we can move on to implementing the next feature.

Table \ref{tab:refactored} shows the runtime of a test problem simulated with the original or refactored solver.
In either case, the solver is compiled with ifort, which is one of the best-performing compilers for BATSRUS on the CPU.
The test problem, which we introduce in detail in section \ref{sec:test_earth}, is to simulate Earth's magnetosphere for the same amount of time.
We run the tests on 64 or 128 cores of one 128-core AMD EPYC 7742 ``Rome'' processor.
Both the original and the refactored codes scale similarly, but the refactored code is noticeably faster, which means that our effort in porting the model to GPUs also improves efficiency on CPUs.

\begin{table}[h]
    \centering
    \caption{CPU runtime comparison: the original versus refactored solvers.}
    \label{tab:refactored}
    \begin{tabular}{ccc}
        \hline
         & Original  & Refactored \\ 
        128-core runtime [s]     &     62.24           &    36.11             \\ 
        64-core runtime [s]    &     120.19         &     71.83         \\ 
        \hline
    \end{tabular}
\end{table}

\subsection{Performance on one GPU}

Following our implementation strategy, we have ported the Geospace application \citep{Pulkkinen:2013} of BATSRUS to a single GPU. The code runs on a single A100 GPU about as fast as on 270 CPU cores, or 2.1 nodes (128-core AMD EPYC 7742 ``Rome'') when the refactored code is compiled with the ifort compiler (the fastest option for CPU execution). This performance is typically the best one can achieve for these types of codes solving partial differential equations on a grid. We did profiling on one V100 GPU and found that the speed is limited by the bandwidth of the main GPU memory. The number of floating-point operations for a given amount of data is too small to make full use of the raw computing power of the GPU.

\section{Porting BATSRUS to Multiple GPUs}
\label{sec:Algorithm}


\subsection{Ghost cell layers}

Each grid block has $n_I\times n_J \times n_K$ physical grid cells. 
To maintain data locality for each block, a layer of ghost cells of width $n_G$ surrounds the physical cells, which store information from neighboring blocks. The total number of grid cells per block is hence $(n_I+2n_G)\times (n_J+2n_G) \times (n_K+2n_G)$. The block size $n_I$, $n_J$, $n_K$, and the ghost cell number $n_G$ are set before compilation. The default value of $n_G$ is 2, corresponding to a second-order accurate scheme. This layer of ghost cells must be filled in every time before the states in physical cells advance in time. Assume, without loss of generality, block B neighbors block A in the $+x$ direction. The ghost cell layer of block A in the $+x$ direction ($x$ indices $n_I+1$ and $n_I+2$) has to be filled with a slice of physical cells from block B, whose cell indices in the $x$ direction are 1 or 2. Note that, in this example, we follow BATL's indexing convention that physical cell indices start from 1.

We intend to run BATSRUS on multiple MPI processes, each of which uses one CPU core and one GPU. When two neighboring blocks are on the same MPI process, filling in ghost cells is realized by local data movements (``local message passing''). When they are on different MPI processes, we use MPI to exchange data among processes (``remote message passing''). 
The following sections introduce the original and new message passing algorithms. The initial data on each process is an array, $\mathrm{State\_VGB}$, which is of size $n_V\times (n_I+2n_G)\times (n_J+2n_G) \times (n_K+2n_G) \times n_B$. 
For clarity, we follow these principles to name the variables in the pseudocode:

\begin{enumerate}
    \item Arrays are in Roman font. Array names consist of a description of the array and its indices, which are separated by an underscore. The number of indices corresponds to the number of variables needed to identify an element of the array. Example: $\mathrm{State\_VGB}$ has three sets of indices -- Variable, Grid, and Block.
    \item Indices and total numbers use subscripts wherever possible. Example: $i_B$ (block index) and $n_B$ (number of blocks).
    \item Intermediate variables in the algorithm use italicized full names. Example: $nSizeS$, $iBufferS$.
    \item For nested loops over $i$, $j$, and $k$, we use the shorthand notation $[ijk]$.
\end{enumerate}


\subsection{Original Message Passing Algorithm}
\label{sec:algo_original}

The original algorithm (Algorithm \ref{algo_cpu}) has been optimized for execution on the CPU. There are 5 steps: Pre-processing the index ranges, determining the sizes of the buffers, filling in the buffers to be sent, exchanging buffers among all processes, and filling in ghost cells with the received buffers. 

There are three types of communications: restriction, where the sending block is finer than its neighbor; prolongation, where the block is coarser; and copy for equal resolution. Each type of communication varies in the cell index ranges to be sent or received, and a pre-processing subroutine obtains the respective ranges, in the form of $[ijk]_{S/R,Min}$ and $[ijk]_{S/R,Max}$, where the $S/R$ in the subscript denotes sending or receiving. For convenience, we also compute $n_C$, which is the additional amount of information we send along with an actual message to provide the receiving block index and cell index ranges. We always have $n_C=1+2n_D$, where $n_D$ is the dimensionality of the problem.

Next, we determine the total size of the information to be sent and received. For each block, we loop over all neighbor directions $i_{Dir}$. By looking up grid information available from BATL (see section~\ref{sec:BATL}), we learn whether the neighboring block in direction $i_{Dir}$ is on a different process and whether it is finer, coarser, or equal in resolution. When there is a pair of messages to be sent to and received from the neighbor in $i_{Dir}$, we add the message sizes to nBufferS\_P and nBufferR\_P, and enlarge the sending and receiving buffers (BufferS\_I and BufferR\_I) by reallocating them. 

To realize single buffer communication, it is necessary to construct an array of indices, iBufferS\_P, whose $i_P$-th element is the starting index of the message sent to process $i_P$. iBufferS\_P is constructed using the sizes of communication to each process, nBufferS\_P.

The serial buffer array, BufferS\_I, stores the messages one after another. Since we have iBufferS\_P, BufferS\_I also stores communication to process $i_P+1$ immediately after that to process $i_P$. To fill in the buffer, we maintain the index $iBufferS$, which denotes the index of $\mathrm{BufferS\_I}$ about to be written with information. In the loops over all blocks ($i_B$), all directions ($i_{Dir}$), and all grid cells to be sent, $iBufferS$ is incremented by $n_V$ for each cell finished. 
Note that we pass the block index and cell index range of the receiving block ($i_B$, $[ijk]_{R,Min}$, $[ijk]_{R,Max}$) along with the actual information from $\mathrm{State\_VGB}$.

We use non-blocking MPI subroutines to send and receive $\mathrm{BufferS\_I}$ and $\mathrm{BufferR\_I}$ when they contain non-zero size data. Using the array storing the starting position (iBufferS\_I), the buffer to be sent (BufferS\_I) is broken into $n_P$ pieces, each of size nBufferS\_P($i_P$), which are then sent to their destination (process $i_P$).
On the receiving side, each process gathers $n_P$ pieces of arrays, each of size nBufferR\_P, into a single receiving buffer BufferR\_I.
All communication is done in parallel to make optimal use of the networking hardware and software.
Once all MPI requests finish for a process, we fill in ghost cells from $\mathrm{BufferR\_I}$ serially. We use the sequentially advanced index $iBufferR$ to retrieve information from the buffer. 
For each message, we first retrieve the block index and index range of the receiving cells ($i_B$, $[ijk]_{R,Min}$, $[ijk]_{R,Max}$), and then fill them with the state information that follows.
In this process, $iBufferR$ advances first by $n_C$ for the receiving block index and cell index ranges, then by $n_V$ for each cell within the index ranges.
We refer readers to \citet{Toth:2012swmf} for more details about the original message passing algorithm.

\subsection{Challenges for Running on Multiple GPUs}
\label{sec_challenge}
The original algorithm does not run efficiently on multiple GPUs. The original algorithm assumes a light workload for message passing for each MPI process, but this is not the case for GPU runs: the number of MPI processes for a GPU run is much smaller than for a CPU run (by a factor of hundreds) if the same speed is desired, so the amount of data to be sent and received by each process is much larger. With the original algorithm, we find that message passing becomes the bottleneck and can take up 90\% of the total run time. In fact, running on 2 GPUs was slower than running on 1. To adapt BATSRUS for a more efficient message passing algorithm on multiple GPUs, we face challenges imposed by (1) the serial bottlenecks in the original algorithm,  (2) the complexity of the communication pattern, and (3) GPU and OpenACC limitations.

Algorithm \ref{algo_cpu} uses $iBufferS$ and $iBufferR$, which advance sequentially cell by cell, to fill in and retrieve data from buffers.
$iBufferS$ and $iBufferR$ for one block has dependence on those for the previous block.
On the other hand, parallelizing the loops over blocks ($i_B$) and/or cells ($[ijk]$) requires $iBufferS$ and $iBufferR$ to be calculated independently for each block and cell.

The communication pattern among all processes is complicated. 
On one hand, message sizes are different. A block can send a message in the direction of its corner, edge, or face. Taking typical values such as $n_I=n_J=n_K=10$, $n_V=8$, and $n_G=2$, a message through a corner is only $n_G^3\times n_V=64$ real numbers, while a message through a face is as large as $n_I\times n_J\times n_G\times n_V= 1,600$ real numbers. 
In the meantime, message sizes can also be affected by resolution changes. If two neighboring blocks are of equal resolution, we simply copy all cells near the boundary. 
If the neighbor of a block is coarser, however, the size of the message is divided by 2 or 4 depending on whether the exchange takes place across an edge or a face. 
On the other hand, the total number of messages is large. In a typical application, there are many thousands of grid blocks on each GPU, and the number of messages sent between any two processes is of a similar order.
The number of messages, combined with the aforementioned size differences, means that simple solutions such as taking the maximum size for each message would waste large amounts of memory. The large number of messages also means that having individual (non-blocking) MPI send/receive requests for each message would be exceedingly inefficient in terms of speed (we actually tried this approach).

GPU acceleration with OpenACC imposes its own limitations.
Moving data between the CPU and GPU is slow, so we must reduce such actions to a minimum. 
Furthermore, we use Fortran allocatable arrays for the buffers, whose sizes are determined during runtime. 
Especially, the buffer sizes change after AMRs.
Since we choose to perform AMRs on the CPU, a part of the algorithm has to run on the CPU to keep track of a possible grid change.
It then must be able to efficiently update the GPU with information that accounts for the change.


\subsection{Message Passing Algorithm for multiple GPUs}


Algorithm \ref{algo_gpu} addresses the issues mentioned above. The pre-processing of the index range remains the same; the counting loop becomes more informative and efficient, and the loops for filling in the buffer and retrieving information become parallel.


We discuss the core ideas of the parallel algorithm before going into its details. We would like to fill in and retrieve from buffers in parallel for all messages, so an MPI process must know the starting index of any message. We use intermediate arrays to store this information, which we call ``memory maps''. Due to the challenges discussed above, two levels of memory maps are needed to facilitate parallelization. 
The first map, denoted by iMsgDir\_IBP, maps $(i_{Dir},iB,i_P)$ to $i_M$ if there is a message for the indices; otherwise, the array element retains the default value of -1, indicating no message for the indices. This array is interpreted as: ``On this process, the message in direction $i_{Dir}$ from block $i_B$ to process $i_P$ has an index of $i_M$.'' The second map, denoted by iBufferS\_IP, maps $(i_M, i_P)$ to $iBuffer$, the index of the starting location of a message. It is interpreted as: ``On this process, the $i_M$-th message to process $i_P$ starts from the $iBuffer$-th real number in the buffer.''
It is possible to directly map $(i_{Dir},i_B,i_P)$ to $iBuffer$, but the resulting array will be inefficient and difficult to interpret from the perspective of the receiving process.
With our two-map strategy, we can construct iBufferS\_IP from the perspective of the sending process and send it to the receiving process, where it automatically becomes the memory map for parallel retrieval.
Lastly, the counting loop has to be run in series because of the challenges discussed in section~\ref{sec_challenge}, so we fill in these two maps on the CPU and send them to the GPU.

To minimize the number of times we move data between the CPU and GPU, we run the counting loop only when the memory maps should change. This can happen if $n_V$ changes (when we reuse the message passing subroutine for other variables than the state) or if the grid changes (after an AMR). We introduce a boolean variable, $IsCounted$, which is only true if all input parameters are the same as the last call. We run the counting loop in series if $IsCounted$ is false. Similar to the CPU algorithm, we loop over $i_B$ and $i_{Dir}$, and maintain $i_M$ to be the index of the newest message. We assign $i_M$ to iMsgDir\_IBP if there is a message, compute the message size $nSizeS$, and store the starting index for $i_M$ in iBufferS\_IP. For every block, we enlarge BufferS\_IP and BufferR\_IP if they are close to being filled up. In the end, we push the updated memory maps to GPUs. 

We can now fill in the buffer in parallel using these memory maps. For every message, we obtain $i_M$ from iMsgDir\_IBP, and map it to $iBufferS$ using iBufferS\_IP. In this way, we have decoupled computation for iBufferS between a message and its predecessor so that the loop over $i_B$ can utilize gang-level parallelization, and the loop over $[ijk]$ can utilize vector-level parallelization.

When exchanging buffers among all processes, we also pass iBufferS\_IP to the receiving processes if $IsCounted$ is false, where they automatically become iBufferR\_IP. The !\$acc host\_data directive makes the addresses of the memory map and the buffer in GPU memory available on the host CPU. When the hardware supports GPU-to-GPU communication, this results in a minimum turnaround time. We then loop over $i_M$ to retrieve from the buffer, and the rest is identical to the original algorithm, except that we no longer keep incrementing $iBufferR$, but rather obtain it from iBufferR\_IP inside parallel loops.

\begin{algorithm}
\label{algo_cpu}
\caption{Original message passing algorithm for CPU.}
\tcc{Pre-process index range.}
Compute $[ijk]_{S,Min}$, $[ijk]_{S,Max}$, $[ijk]_{R,Min}$, $[ijk]_{R,Max}$ for restriction, prolongation, and copy for equal resolution\;
Compute the number of indices needed to store the index range of a message, $n_C\leftarrow 1+2n_D$\;
\tcc{Set the size of buffers sent to and received from other processes.}
$\mathrm{nBufferS\_P}\leftarrow 0$; $\mathrm{nBufferR\_P \leftarrow 0}$\;
\For{$i_B=1,n_B$}{
\For{$i_{Dir}=1,n_{Dir}$}{
\If{
$\mathrm{There~is~a~message~to~process~}i_P\neq\mathrm{self}$
}{
Determine if the neighboring block in direction $i_{Dir}$ has finer, equal, or coarser resolution\;
Compute sizes of sent and received messages: $nSizeS$, $nSizeR$\;
$\mathrm{nBufferS\_P(i_P)}\leftarrow \mathrm{nBufferS\_P(i_P)}+nSizeS$\;
$\mathrm{nBufferR\_P(i_P)}\leftarrow \mathrm{nBufferR\_P(i_P)}+nSizeR$\;
}
}
}
If needed (re)allocate(BufferS\_I(sum(nBufferS\_P)), BufferR\_I(sum(nBufferR\_P)))\;
\tcc{Set the start indexes in the send buffer for each prorcessor, iBufferS\_P.}
\For{$i_P=1,n_P$}{
iBufferS\_P($i_P$) $\leftarrow$ sum(nBufferS\_P(1:$i_P-1$))\;
}
\tcc{Fill BufferS\_P with index range and data from physical (non-ghost) grid cells.}
\For{$i_B=1,n_B$}{
\For{$i_{Dir}=1,n_{Dir}$}{
\If{$\mathrm{There~is~a~message~to~process~}i_P$}{
\If{$i_P=\mathrm{self}$}{ 
Local message passing\;
}
\Else{
Determine if restriction, prolongation or copy for equal resolution takes place\;
$iBufferS\leftarrow$iBufferS\_P($i_P$)\;
$\mathrm{BufferS\_I}(iBufferS+1:iBufferS+n_C)\leftarrow i_B,[ijk]_{R,Min}, [ijk]_{R,Max}$\;
$iBufferS\leftarrow iBufferS+n_C$\;
\For{$[ijk]=[ijk]_{S,Min}, [ijk]_{S,Max}$}{
$\mathrm{BufferS\_I}(iBufferS+1:iBufferS+n_V)\leftarrow\mathrm{State\_VGB}(:,i,j,k,i_B)$\;
$iBufferS\leftarrow iBufferS+n_V$\;
}
}
}
}
}
\tcc{Exchange buffers.}
Send (non-blocking) $i_P$-th segment of $\mathrm{BufferS\_I}$ to process $i_P$\;
Receive (non-blocking)  $i_P$-th segment of $\mathrm{BufferR\_I}$ from process $i_P$\;
Wait for all sending and receiving to finish\;

\tcc{Fill in ghost cells from BufferR\_I.}
$iBufferR \leftarrow 0$\;
\While{$iBufferR<\mathrm{sum(nBufferR\_P)}$}{
\tcc{Get receiving block index and grid cell index range.}
$(i_B,[ijk]_{R,Min},[ijk]_{R,Max})\leftarrow \mathrm{BufferR\_I}(iBufferR+1:iBufferR+n_C)$\;
$iBufferR\leftarrow iBufferR+n_C$\;
\For{$[ijk]=[ijk]_{R,Min}, [ijk]_{R,Max}$}{
$\mathrm{State\_VGB}(:,i,j,k,i_B)\leftarrow\mathrm{BufferR\_I}(iBufferR+1:iBufferR+n_V)$\;
$iBufferR\leftarrow iBufferR + n_V$\;
}
}
\end{algorithm}

\begin{algorithm}
\label{algo_gpu}
\caption{Parallel message passing algorithm for GPU.}
\tcc{Pre-process index range. This part is unchanged}
Compute $[ijk]_{{S,Min}}$, $[ijk]_{{S,Max}}$, $[ijk]_{{R,Min}}$, $[ijk]_{{R,Max}}$ for restriction, prolongation, and copy for equal resolution\;
Compute the number of indices needed to store the index range of a message, $n_C\leftarrow 1+2n_D$\;
$IsCounted\leftarrow$``If input parameters are unchanged"\;
\If{$\mathrm{not}~IsCounted$}{
\tcc{Calculate the buffer sizes and construct memory maps on the CPU.}
Initialize $\mathrm{nMsg\_P}$, $\mathrm{iMsgDir\_IBP}$, and $\mathrm{iBufferS\_IP}$\;
\For{$i_B=1,n_B$}{
\For{$i_{Dir} = 1, n_{Dir}$}{
\If{$\mathrm{there~is~a~message~in~} i_{Dir}$}{
Look up the destination process $i_P$ using grid information\;
\If{$i_P\neq\mathrm{self}$}{ 
$i_M\leftarrow \mathrm{nMsg\_P}(i_P)$\;
$\mathrm{iMsgDir\_IBP}(i_{Dir},i_B,i_P)\leftarrow i_M$\;
Compute message size, $nSizeS$, depending on message type\;
$\mathrm{iBufferS\_IP}(i_M,i_P)\leftarrow \mathrm{iBufferS\_IP}(i_M-1,i_P)+nSizeS$\;
$\mathrm{nMsg\_P}(i_P)\leftarrow i_M+1$\;
}
}
}
Enlarge $\mathrm{BufferS\_IP}$, $\mathrm{BufferR\_IP}$ as needed\;
}
\tcc{Copy $\mathrm{iMsgDir\_IBP}, \mathrm{iBufferS\_IP}$ to the GPU}
\textcolor{red}{!\$acc update device($\mathrm{iMsgDir\_IBP}$, $\mathrm{iBufferS\_IP}$)}
}
\tcc{Fill in $\mathrm{BufferS\_IP}$ on the GPU.}
\textcolor{red}{!\$acc loop gang private($i_M$, $i_P$, $iBuffer$)}\\
\For{$i_B=1,n_B$}{
\For{$i_{Dir}$ = 1, $n_{Dir}$}{
\If{$\mathrm{there~is~a~message~in~direction~} i_{Dir}$}{
Look up the destination process $i_P$\;
\uIf{$i_P=\mathrm{self}$}{
Local message passing\;
}
\Else{ 
$i_{B,R}\leftarrow$ Receiving block index on $i_P$\;
$i_M\leftarrow \mathrm{iMsgDir\_IBP}(i_{Dir}, i_B, i_P)$\;
$iBuffer\leftarrow \mathrm{iBufferS\_IP}(i_M,i_P)$\;
$\mathrm{BufferS\_IP}(iBuffer+1:iBuffer+n_C) \leftarrow i_{B,R}, [ijk]_{R,Min}, [ijk]_{R,Max}$\;

\textcolor{red}{!\$acc loop vector collapse(4) private($iBufferS$)}\\
\For{$i_V=1,n_V;[ijk]=[ijk]_{S,Min}, [ijk]_{S,Max}$}{
$iBufferS \leftarrow iBuffer,i_V,i,j,k$\;
$\mathrm{BufferS\_IP}(iBufferS)\leftarrow\mathrm{State\_VGB}(i_V,i,j,k,i_B)$\;
}
}
}
}
}
\tcc{Exchange buffers and memory maps using addresses on the GPU.}
\textcolor{red}{!\$acc host\_data use\_device(BufferS\_IP, iBufferS\_IP,BufferR\_IP,iBufferR\_IP)}\\
Send $\mathrm{BufferS\_IP}$ (and $\mathrm{iBufferS\_IP}$ if not $IsCounted$) to all other processes\;
Receive $\mathrm{BufferR\_IP}$ (and $\mathrm{iBufferR\_IP}$ if not $IsCounted$) from all other processes\;
Wait for all sending and receiving to finish\;
%
\tcc{Fill in ghost cells on the GPU.}
\For{$i_P\neq\mathrm{self}$}{
\textcolor{red}{!\$acc loop gang private($iBuffer$)}\\
\For{$i_M=1,\mathrm{nMsg\_P}(i_P)$}{
$iBuffer\leftarrow \mathrm{iBufferR\_IP}(i_M,i_P)$\;
$i_B,[ijk]_{R,Min},[ijk]_{R,Max}\leftarrow \mathrm{BufferR\_IP}(iBuffer+1:iBuffer+n_C,i_P)$\;
\textcolor{red}{!\$acc loop vector collapse(4) private($iBufferR$)}\\
\For{$i_V=1,n_V;[ijk]=[ijk]_{R,Min}, [ijk]_{R,Max}$}{
$iBufferR \leftarrow iBuffer,i_V,i,j,k$\;
$\mathrm{State\_VGB}(i_V,i,j,k,i_B)\leftarrow\mathrm{BufferR\_IP}(iBufferR,i_P)$\;
}
}
}
\end{algorithm}

\section{Test and Performance}
\label{sec:tests}

During the course of development for the GPU implementation, we perform daily regression testing using various test cases, from a simple 1-D shock tube test to the evolution of the solar corona using the Alfv\'{e}n Wave Solar Model (AWSoM) \citep{vanderHolst:2010}. We run the tests on 1 to 3 GPUs to test the new message passing algorithm, where the 3-GPU runs are intended to reveal bugs related to load balancing. Among these tests, we select the fast wave test to demonstrate weak and strong scaling of the GPU code. Later, we perform an Earth magnetosphere simulation to demonstrate the advantages of using multiple GPUs. 

\subsection{Scaling Tests with a 3D MHD Problem}
The fast wave test solves the ideal MHD equations in nearly conservative form:
\begin{eqnarray}
    \frac{\partial\rho}{\partial t} &+& \nabla\cdot({\rho \mathbf u}) = 0 
    \label{eq:drhodt} \\
    \frac{\partial\rho\mathbf u}{\partial t} &+& \nabla\cdot\left[{\rho \mathbf u \mathbf u}
    -\mathbf B\mathbf B + 
    I \left(p + \frac{B^2}{2}\right)\right] = -\mathbf B  \nabla \cdot \mathbf B 
    \label{eq:drhoudt} \\
        \frac{\partial \mathbf B}{\partial t} &+& \nabla \cdot (\mathbf u \mathbf B - \mathbf B \mathbf u) = -\mathbf u \nabla \cdot \mathbf{B}
    \label{eq:dbdt}\\
    \frac{\partial e}{\partial t} &+& \nabla\cdot\left[\mathbf u\left(e + p + \frac{1}{2}B^2\right) -\mathbf u \cdot \mathbf B \mathbf B\right] = -\mathbf u \cdot \mathbf B \nabla \cdot \mathbf B,
    \label{eq:dedt}
\end{eqnarray}
where the total energy density 
\begin{eqnarray}
\label{eq:energy}
e = \frac{p}{\gamma-1} + \frac{\rho u ^2}{2} + \frac{B^2}{2}.
\end{eqnarray}
The source terms proportional to $\nabla \cdot \mathbf{B}$ were introduced to control the numerical error in $\nabla \cdot \mathbf{B}$, which was proposed as the eight-wave scheme by \citet{Powell:1999}. The units of the magnetic field are chosen to make the vacuum permeability $\mu_0=1$.

The setup for numerical simulation is as follows. 
We first divide a 3-D computational domain uniformly into $N_x\times N_y \times N_z$ grid blocks. In each grid block, $n_I=n_J=n_K=10$, so there are 1000 grid cells per block. The domain is the space bounded by $[-32, 32]\times[-32, 32]\times[-32,32]$. 
One-eighth of the domain in the corner, which is bounded by $[-32, 0]\times[-32, 0]\times[-32,0]$ is then refined by a factor of 2 to test our algorithm's ability to deal with a change in grid resolution. 
Furthermore, we use an exponent $m$ to denote various grid sizes per process.
The size of the grid can change in two ways:
\begin{enumerate}
\item When $m$ is fixed, the total number of root blocks ($N_xN_yN_z$) is proportional to the number of processes ($n_P$). Since the AMR region stays the same, the total number of blocks after AMR ($n_B$) is also proportional to $n_P$. This corresponds to constant workload per process after load balancing.
\item When the number of processes ($n_P$) is fixed, $N_x=2^\frac{m}{3}N_{x0}$, $N_y=2^\frac{m}{3}N_{y0}$ and $N_z=2^\frac{m}{3}N_{z0}$ for $m\in\{0,3,6,9,12\}$. $N_{x0}$, $N_{y0}$ and $N_{z0}$ are fixed for each $n_P$. This ensures all grids for various values of $m$ are self-similar for a certain $n_P$.
\end{enumerate}
The parameters used to construct the computational grid for any $m$ are shown in Table \ref{tab:parameters_fastwave}.
\begin{table}[h!]
\centering
 \caption{Size of the computational grid for the fast wave tests as functions of $m$.}
 \label{tab:parameters_fastwave}
    \begin{tabular}{lccccccccc} 
 \tableline
Number of GPUs & 1 & 2 & 4 & 8 & 16 & 32 & 64 & 128 & 256 \\
Number of root blocks along $X$: $N_x$ & 2$\cdot 2^\frac{m}{3}$ & 4$\cdot 2^\frac{m}{3}$ & 4$\cdot 2^\frac{m}{3}$ & 4$\cdot 2^\frac{m}{3}$ & 8$\cdot 2^\frac{m}{3}$ & 8$\cdot 2^\frac{m}{3}$ & 8$\cdot 2^\frac{m}{3}$ & 16$\cdot 2^\frac{m}{3}$ & 16$\cdot 2^\frac{m}{3}$ \\
Number of root blocks along $Y$: $N_y$  & 2$\cdot 2^\frac{m}{3}$ & 2$\cdot 2^\frac{m}{3}$ & 4$\cdot 2^\frac{m}{3}$ & 4$\cdot 2^\frac{m}{3}$ & 4$\cdot 2^\frac{m}{3}$ & 8$\cdot 2^\frac{m}{3}$ & 8$\cdot 2^\frac{m}{3}$ & 8$\cdot 2^\frac{m}{3}$ & 16$\cdot 2^\frac{m}{3}$ \\
Number of root blocks along $Z$: $N_z$  & 2$\cdot 2^\frac{m}{3}$ & 2$\cdot 2^\frac{m}{3}$ & 2$\cdot 2^\frac{m}{3}$ & 4$\cdot 2^\frac{m}{3}$ & 4$\cdot 2^\frac{m}{3}$ & 4$\cdot 2^\frac{m}{3}$ & 8$\cdot 2^\frac{m}{3}$ & 8$\cdot 2^\frac{m}{3}$ & 8$\cdot 2^\frac{m}{3}$ \\
Total number of root blocks: 
$N_x N_y N_z$ & 8$\cdot 2^m$ & 16$\cdot 2^m$ & 32$\cdot 2^m$ & 64$\cdot 2^m$ & 128$\cdot 2^m$ & 256$\cdot 2^m$ & 512$\cdot 2^m$ & 1024$\cdot 2^m$ & 2048$\cdot 2^m$ \\
Total number of blocks after AMR: $n_B$ & 15$\cdot 2^m$ & 30$\cdot 2^m$ & 60$\cdot 2^m$ & 120$\cdot 2^m$ & 240$\cdot 2^m$ & 480$\cdot 2^m$ & 960$\cdot 2^m$& 1920$\cdot 2^m$& 3840$\cdot 2^m$ \\
\tableline
 \end{tabular}

\end{table}

The initial state is a fast magnetosonic wave propagating in the $x$ direction defined as
\begin{equation}
\label{eqn_wave}
\mathbf{V}(\mathbf{x})=\mathbf{V}_0 + \mathbf{V_1}\cos(\mathbf{k}\cdot \mathbf{x}),
\end{equation}
where $\mathbf{V}(x)$ is the local state vector, $\mathbf{V}_0$ is the uniform background, $\mathbf{V_1}$ is the amplitude of the wave, and $\mathbf{k}$ is the wave vector. We set $\mathbf{k}$ perpendicular to $\mathbf{B_0}$, so the fast mode propagates at a velocity $c_F=\sqrt{c_s^2+v_A^2}$, where $c_s$ and $v_A$ are the speeds of sound and Alfv\'{e}n waves, respectively. The values of the parameters are:
\begin{gather*}
\mathbf{V_0}=
\begin{bmatrix}
\rho_0 \\ u_{x0} \\ u_{y0} \\ u_{z0} \\ B_{x0} \\ B_{y0} \\ B_{z0} \\ p_0
\end{bmatrix}=
\begin{bmatrix}
1.0 \\ 0.0 \\ 0.0 \\ 0.005 \\ 0.0 \\ 0.04 \\ 0.0 \\ 0.00054
\end{bmatrix},\qquad
\mathbf{V_1}=\begin{bmatrix}
\rho_1 \\ u_{x1} \\ u_{y1} \\ u_{z1} \\ B_{x1} \\ B_{y1} \\ B_{z1} \\ p_1
\end{bmatrix}=
\begin{bmatrix}
0.1 \\ 0.005 \\ 0.0 \\ 0.0 \\ 0.0 \\ 0.004 \\ 0.0 \\ 0.00009
\end{bmatrix},\qquad
\mathbf{k}=\begin{bmatrix}
\frac{\pi}{16}, \\ 0 \\0 
\end{bmatrix}, \qquad 
c_A = 0.04,\ \ c_s = 0.03,\ \ c_F = 0.05.
\end{gather*}
 a second second-order scheme with the Linde flux \citep{Linde:2002_solver} and a CFL number of 0.8 for the simulation, and we impose periodic boundary conditions. 
The simulations are run for a fixed 500 steps so that the total work per process remains the same for various numbers of GPUs.

\begin{figure}
\gridline{\fig{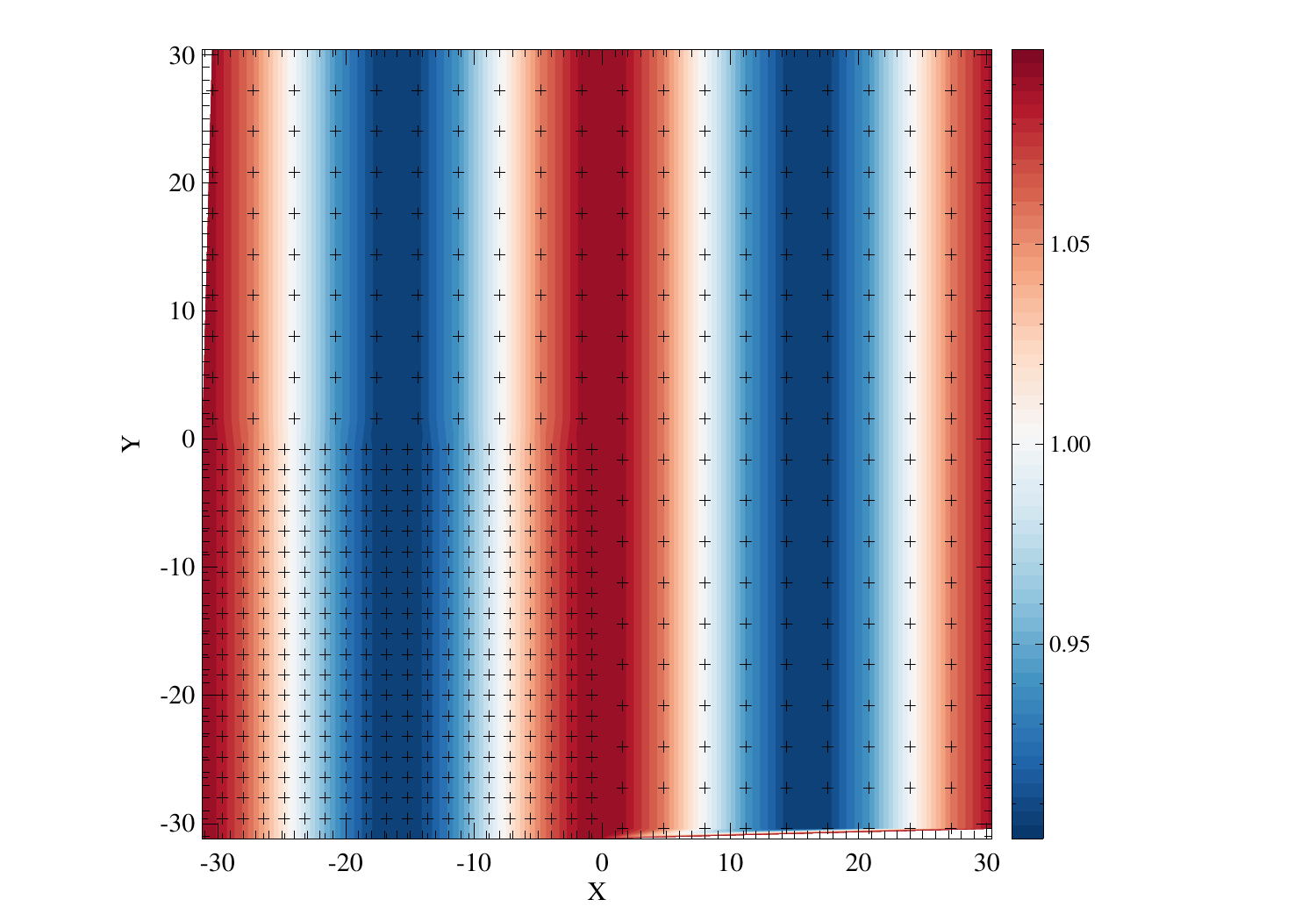}{0.5\textwidth}{(a)}
          \fig{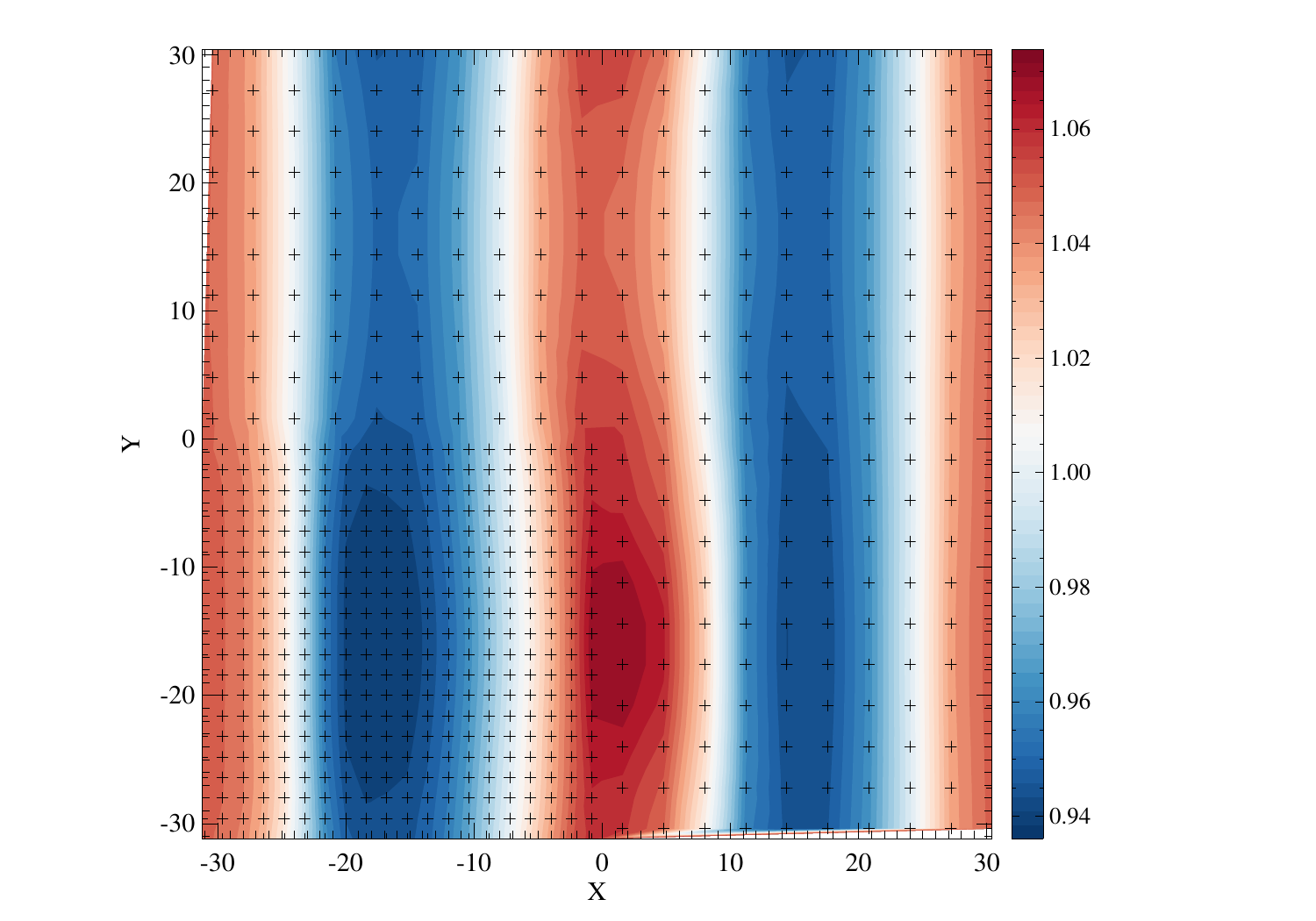}{0.5\textwidth}{(b)}}
\gridline{\fig{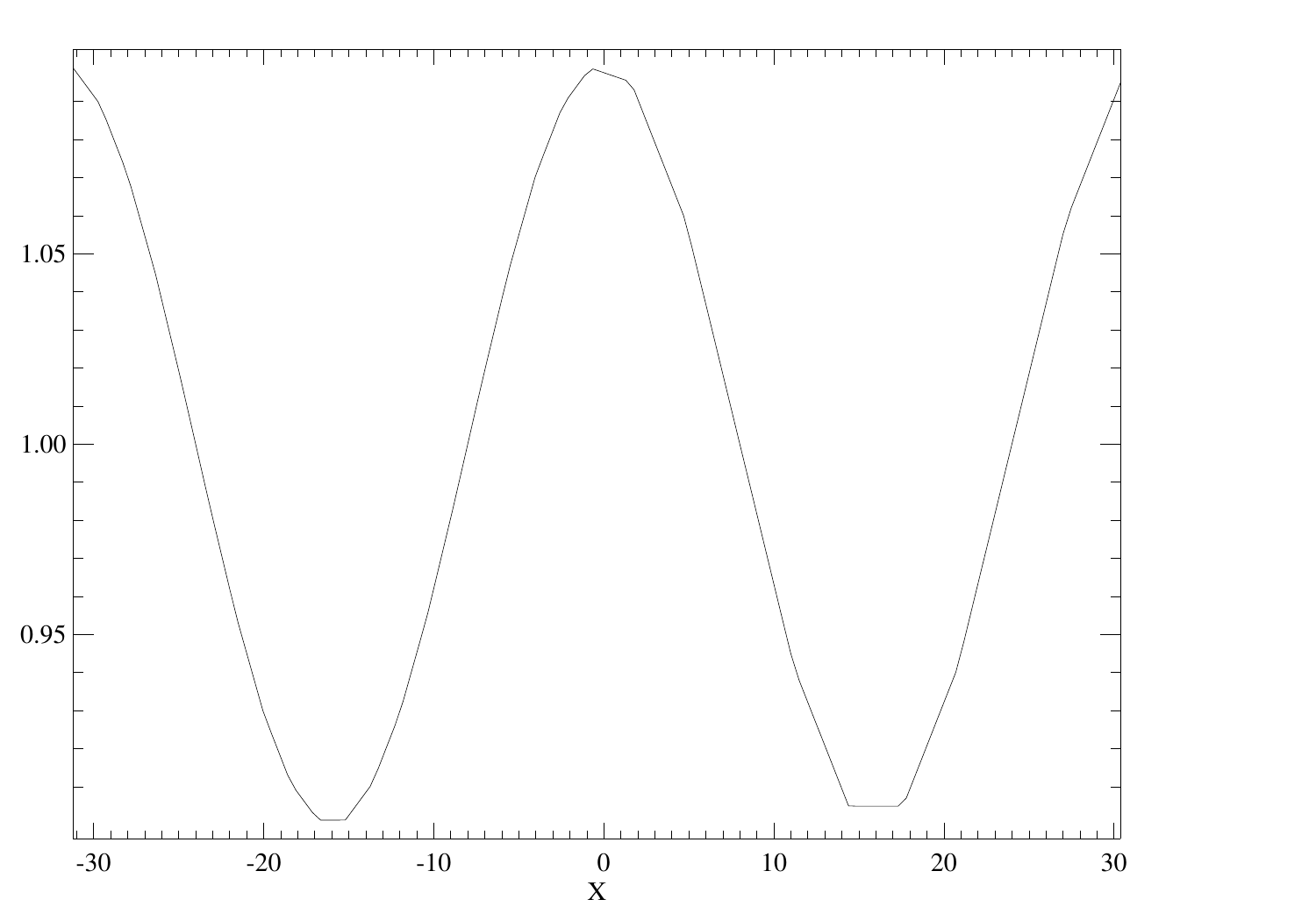}{0.5\textwidth}{(c)}
        \fig{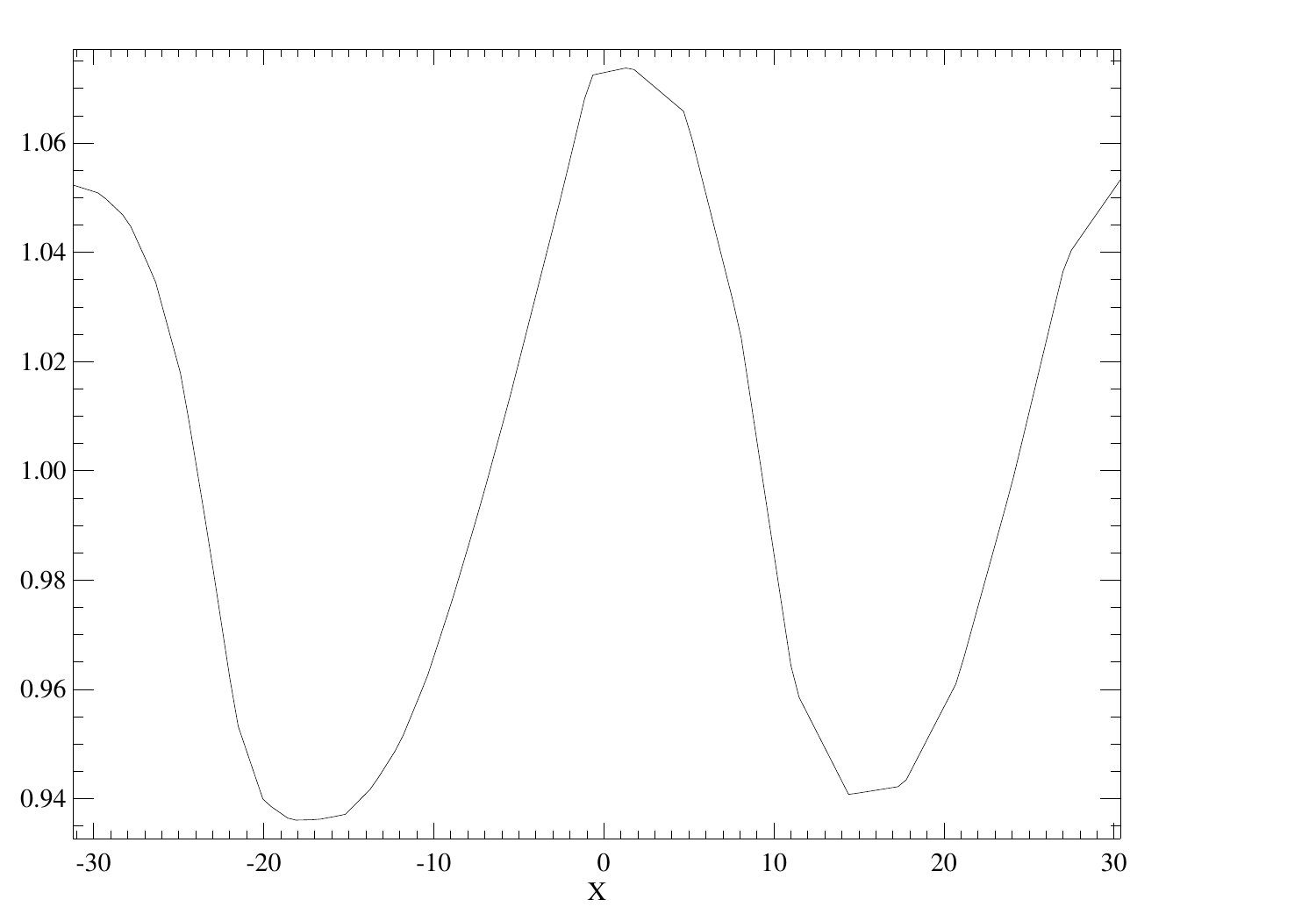}{0.5\textwidth}{(d)}}
\caption{$\rho$ at the start and end of one period of the fast wave. (a) and (b): initial and final state in the $z=-16$ plane; (c) and (d): initial and final state along the $y=-16$, $z=-16$ cut. Note the grid refinement, the effects of resolution change, and slope steepening after one period.}
\label{fig:setup_fw}
\end{figure}

\begin{figure}
\gridline{\fig{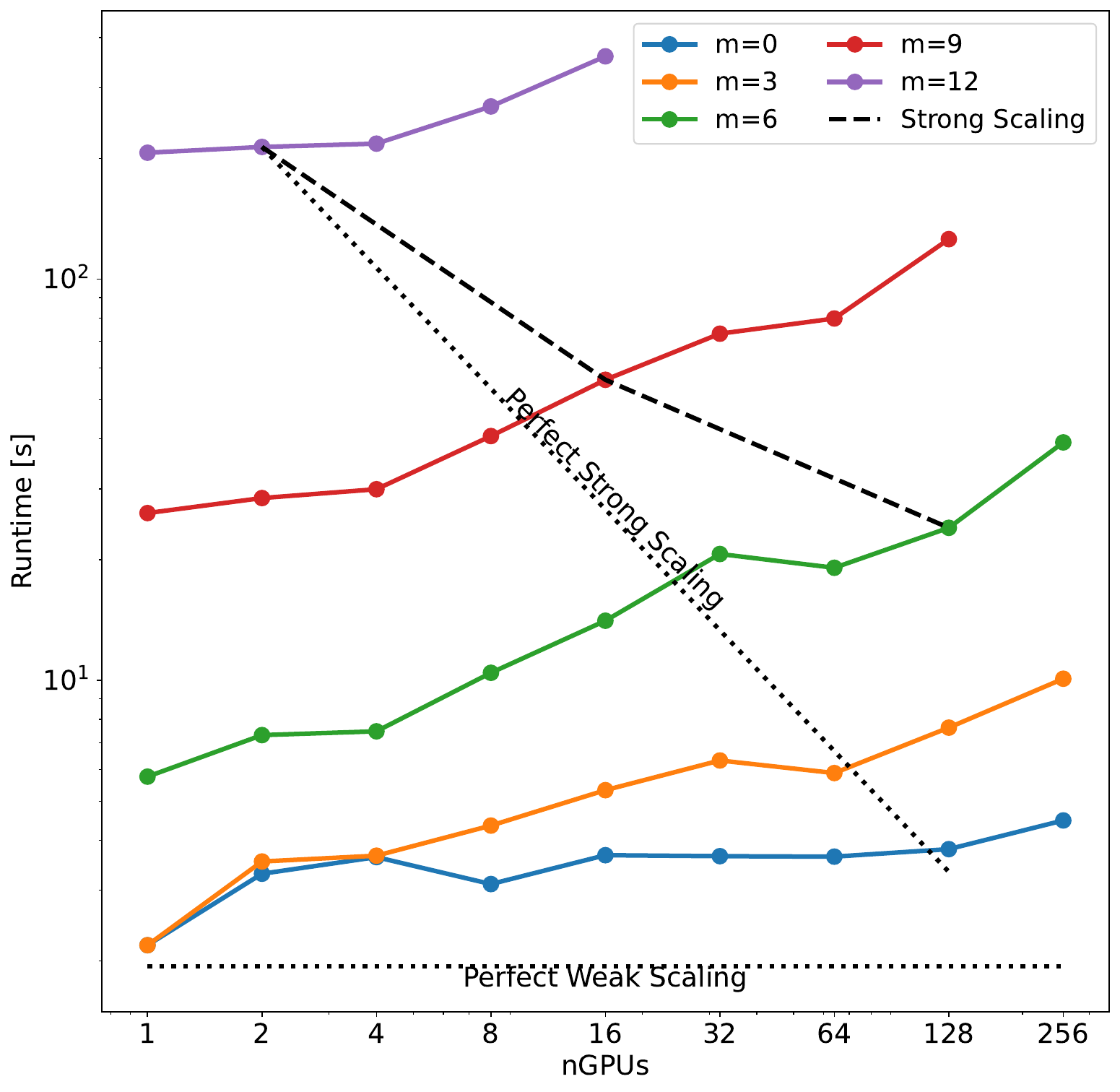}{0.5\textwidth}{(a)}
          \fig{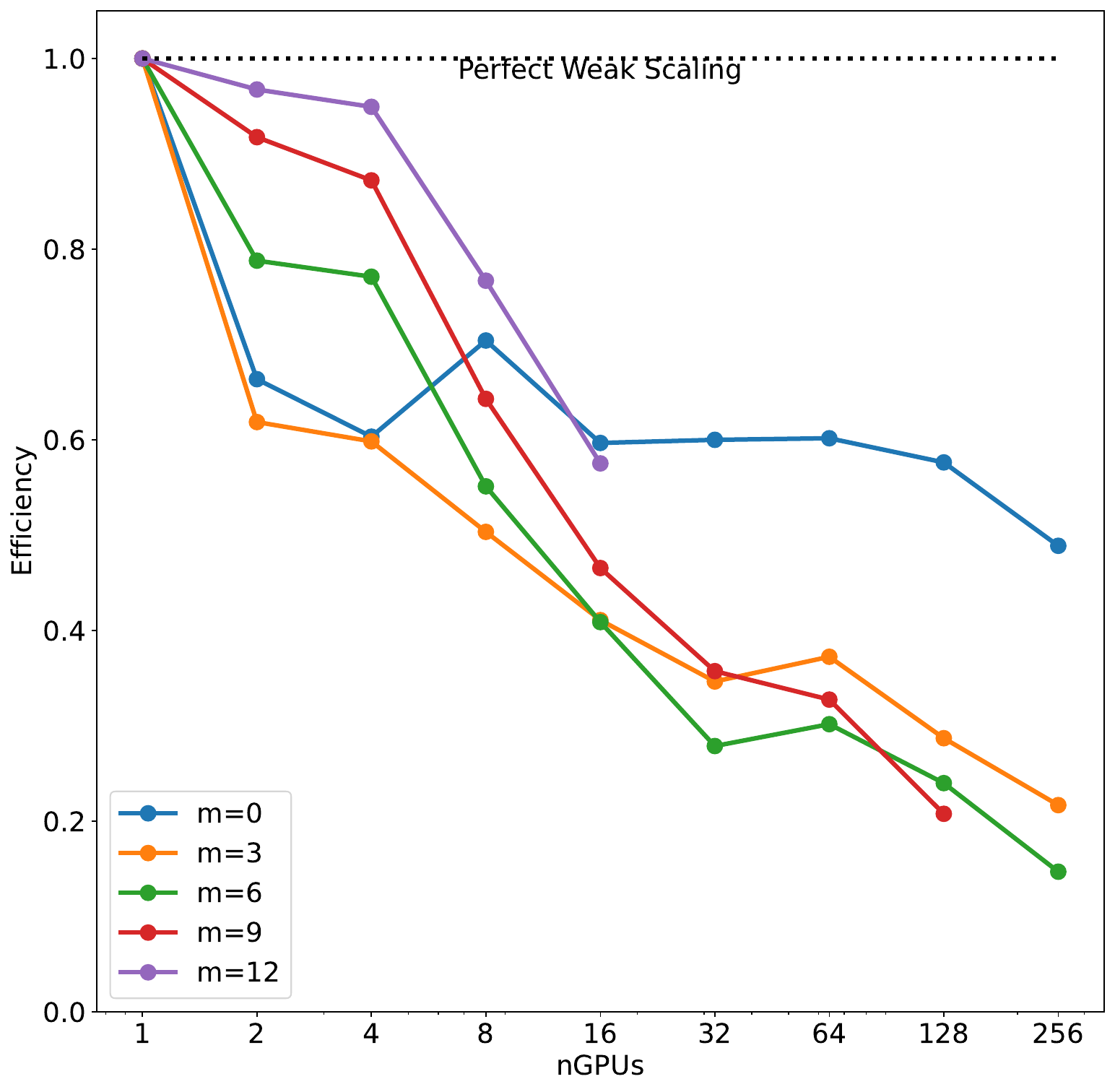}{0.5\textwidth}{(b)}}
\gridline{\fig{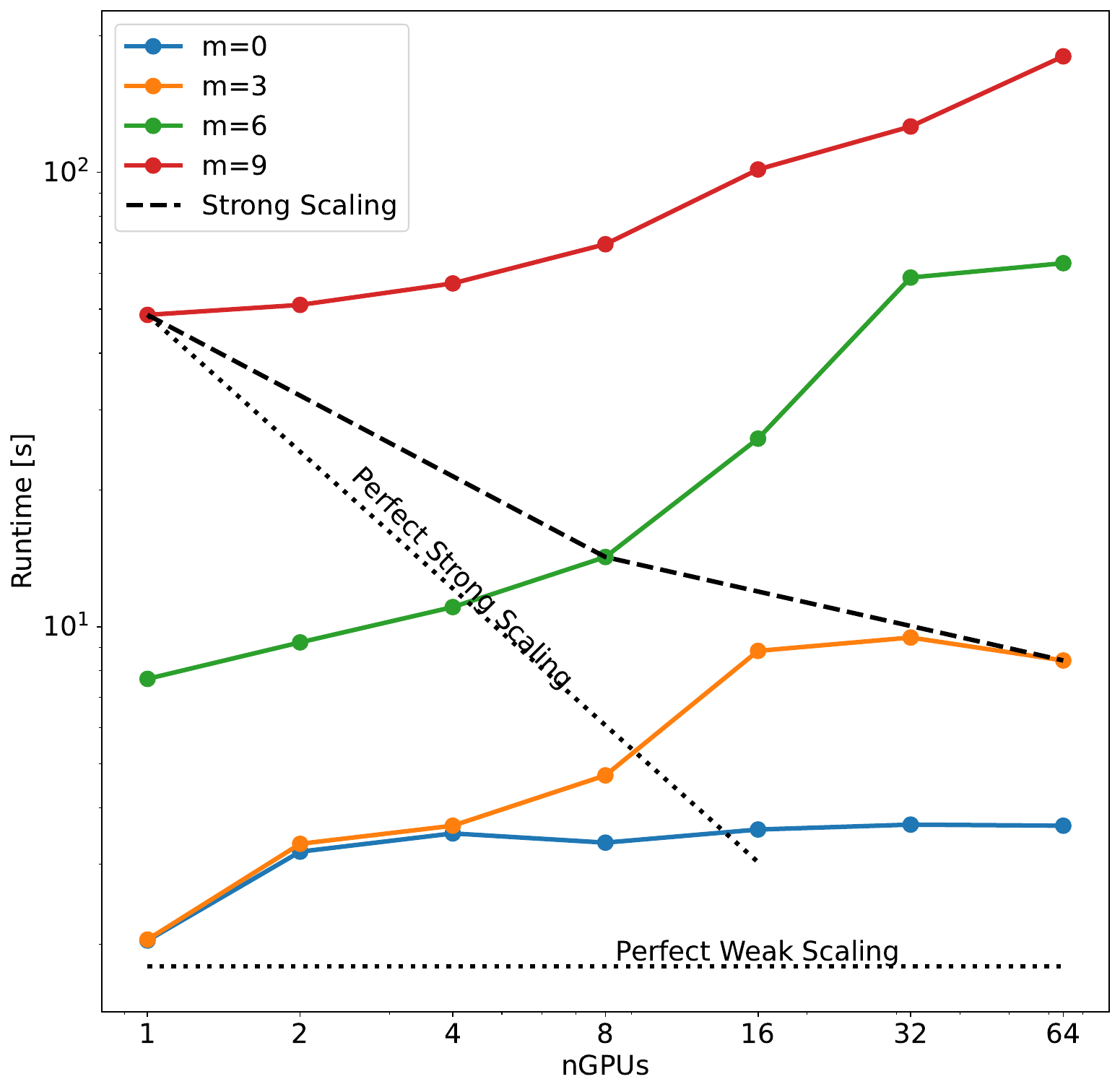}{0.5\textwidth}{(c)}
        \fig{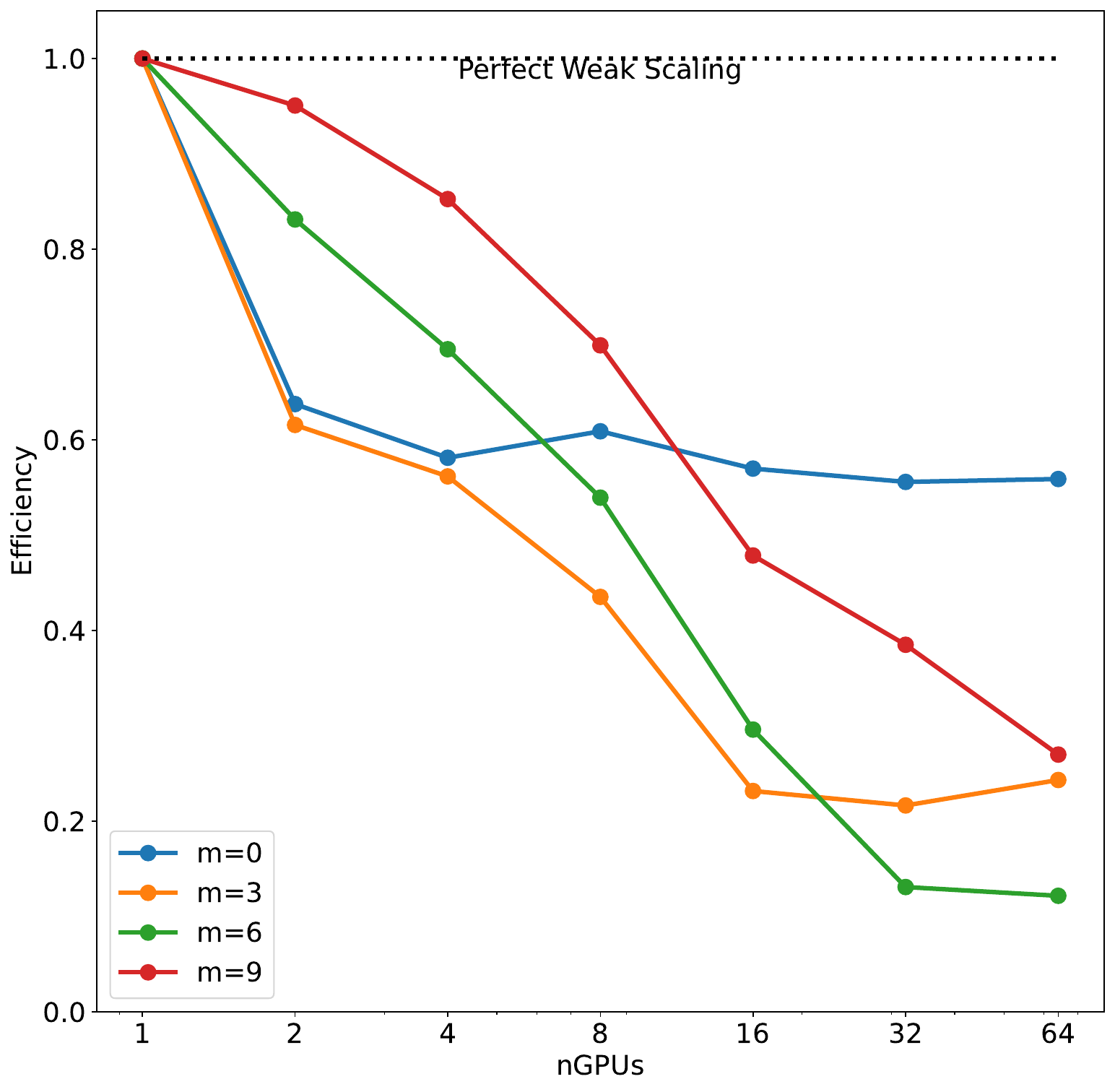}{0.5\textwidth}{(d)}}
\caption{Timings and parallel efficiency from scaling tests. (a)(b) A100 GPUs; (c)(d) V100 GPUs.}
\label{fig:scaling}
\end{figure}

We run scaling tests on the Pleiades supercomputer maintained by the NASA Ames Center. 
Two types of GPUs are available on Pleiades: NVIDIA A100 (80 GB) and V100 (32 GB).
The A100 GPU nodes support 600 GB/s total bandwidth for GPU-to-GPU communication, which takes place exclusively within a node.
The GPUs are connected to the CPU socket via a 31.5 GB/s PCIe 4.0 bus. 
The nodes are connected with High Data Rate (HDR) devices and switches, which provide a maximum of 25 GB/s bandwidth between nodes.
There are four A100 GPUs per node, and we can access at most 64 nodes at a time.
For V100 GPU nodes, GPU-to-GPU communication bandwidth is 300 GB/s, GPU-to-CPU bandwidth is 15.75 GB/s, and maximum node-to-node bandwidth is 12.5 GB/s. Neither type of node supports GPU-to-GPU communication between different nodes.
There are four V100 GPUs per node, while the maximum number of nodes available is 16.

The results are shown in Figure \ref{fig:scaling}. 
Each color represents a weak scaling curve (fixed $m$ or constant problem size per process), while dashed lines show representative strong scaling curves (fixed $n_B$ or constant total problem size). 
The parallel efficiency, denoted by $\eta$, is computed for each fixed $m$:
\begin{equation}
    \label{eqn_efficiency}
    \eta = \frac{t_\text{single-GPU}}{t_\text{multi-GPU}}.
\end{equation}
Note that the single-GPU code is naturally faster than the multi-GPU code, even if the problem size per process remains constant. We specially optimize the single-GPU code. For example, by removing optional subroutine inputs that are only needed for multi-GPU communication, we improve the speed of the single-GPU code substantially. The single-GPU code is also shorter since it skips anything related to the use of buffers. 

\begin{figure}
\centering
\includegraphics[width=\linewidth]{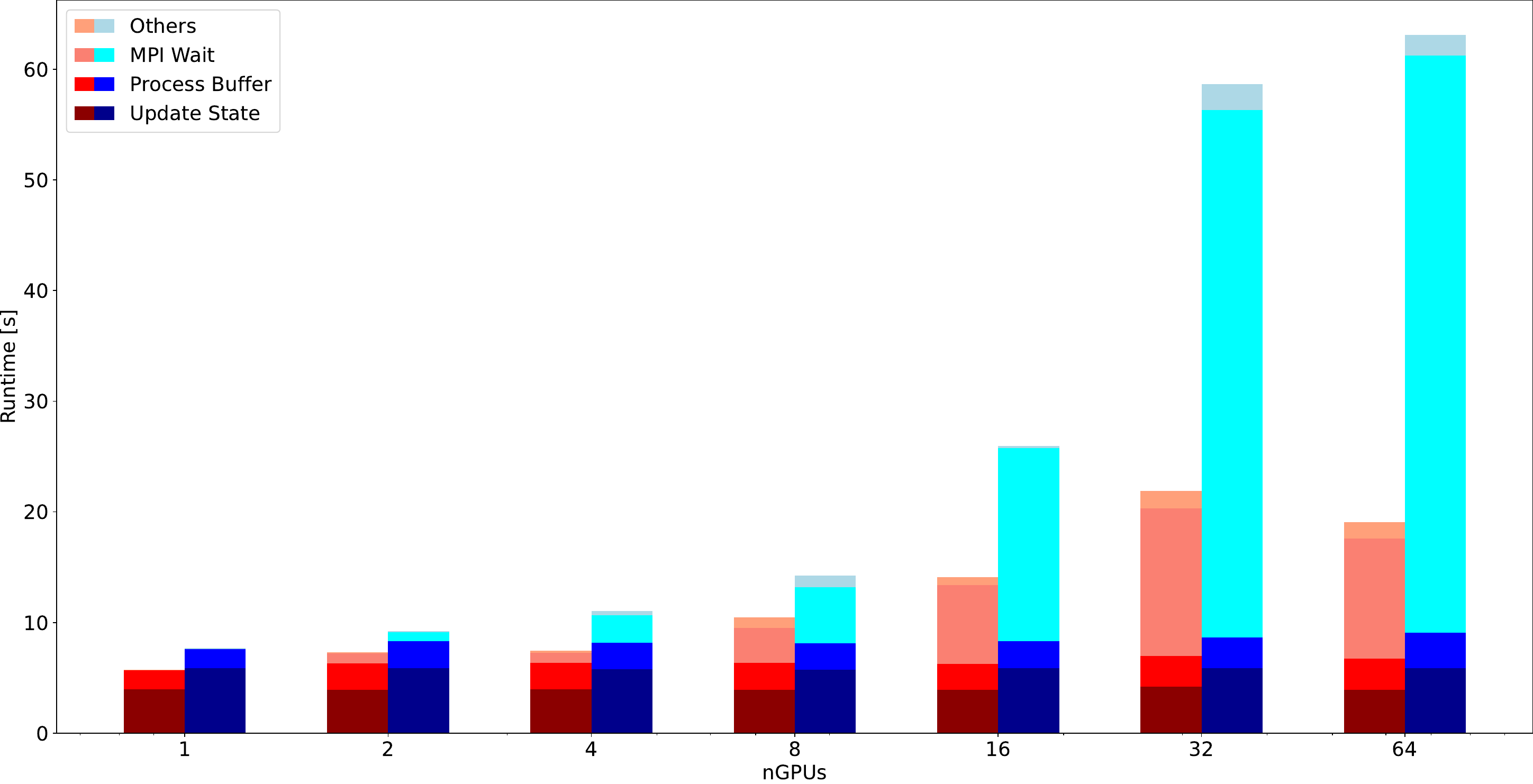}
\caption{Component-wise timings for A100 and V100 GPUs. Red bars are for A100 GPUs, while Blue ones are for V100 GPUs.}
\label{fig:gpu_small}
\end{figure}

The $m=0$ curves show good weak scalability. 
On both types of GPUs, the parallel efficiency drops to around 60\% and 50\% for 64 V100 GPUs and 256 A100 GPUs, respectively.
This shows that the GPU code is algorithmically efficient. 
For larger $m$ values, weak scalability remains good for up to 4 GPUs. 
For $m=12$, as the GPUs are saturated with work, $\eta$ can be as high as around 95\% for 2 or 4 GPUs.
However, it declines for over 4 GPUs and drops to around 20\% for A100 GPUs and less than 20\% for V100. 
This is due to congestion in the network and GPU-to-CPU transfer: for 2 GPUs on different nodes to communicate, data have to be sent to and received from the CPUs first. 
Moreover, the GPU-to-GPU bandwidth is 24 times the node-to-node bandwidth and 19 times the GPU-to-CPU bandwidth; hence, as the amount of data transfer increases with $m$, message passing still finishes fast within one node but fails to keep up for more than one node. 
Figure \ref{fig:gpu_small} and Table \ref{tab:gpu_small} show timings for each component for $m=6$ runs.
``Process buffer'' and ``Update State'', and other tasks that do not involve communication run in almost constant time. 
The increase in runtime for large numbers of GPUs is solely due to the time it takes to finish MPI communications among nodes.

There are other minor findings. 
Firstly, it takes a sufficiently large problem to saturate a GPU.
For $m=0$ or $m=3$, the curves are close together for both GPUs, which means the parallelism the GPUs offer is underutilized. 
The curves start to separate from $m=6$ (960 blocks per process), a sign of ``reasonable'' GPU utilization.
Secondly, for fixed $m$ and $n_P$, the runtime still depends on the number of nodes or the number of GPUs per node.
This can explain some reversals in runtime in the curves, but as many hardware bottlenecks come into play here, we refrain from attempting to interpret them.
Thirdly, while $\eta$ is low with ``reasonable'' GPU utilization, running a large problem on many GPUs is still practical:
increasing the number of GPUs improves the runtime with diminishing returns, and a large problem may simply not fit on one GPU due to memory restrictions.
The largest problem in Figure \ref{fig:scaling} is $m=9$, $n_P=128$ (also $m=12$, $n_P=16$): it has 983 million grid cells!

\begin{table}[]
    \centering
    \caption{Timings for various tasks on A100 and V100 GPUs.}
    \label{tab:gpu_small}
    \begin{tabular}{llccccccc}
    \tableline
    \multirow{5}{20mm}{A100} & nGPUs & 1 & 2 & 4 & 8 & 16 & 32 & 64 \\
    & Total & 5.76 & 7.31 & 7.47 & 10.45 & 14.09 & 21.88 & 19.09 \\
    & MPI wait & n/a & 0.88& 0.91& 3.15& 7.12& 13.35& 10.84
    \\
    & Process Buffer & 1.75 & 2.41 & 2.39 & 2.42 & 2.37 & 2.76 & 2.82 \\
    & Update State & 3.95 & 3.92 & 3.95& 3.92& 3.91&4.21&3.92\\
    \tableline
    \multirow{5}{20mm}{V100} & nGPUs & 1 & 2 & 4 & 8 & 16 & 32 & 64 \\
    & Total & 7.68 & 9.24 & 11.05 & 14.24 & 25.94 & 58.67 & 63.10 \\
    & MPI wait & n/a & 0.82 & 2.50 & 5.04 & 17.48 & 47.63 & 52.13 \\
    & Process Buffer & 1.72 & 2.43 & 2.38 & 2.41 & 2.40 & 2.80 & 3.22 \\    
    & Update State & 5.90 & 5.88 & 5.78 & 5.72 & 5.90 & 5.87 & 5.88\\
    \tableline
    \end{tabular}
\end{table}

\subsection{Earth Magnetosphere Application}
\label{sec:test_earth}
Here we showcase the ability of GPU-accelerated BATSRUS to run space weather forecasts faster than real time. We run BATSRUS with local time stepping to obtain a (quasi-)steady state solution of Earth's magnetosphere. From the steady state magnetosphere, we start a second session in time-accurate mode and compare the run time with the simulated time. 

\begin{figure}
    \centering
    \includegraphics[width=\linewidth]{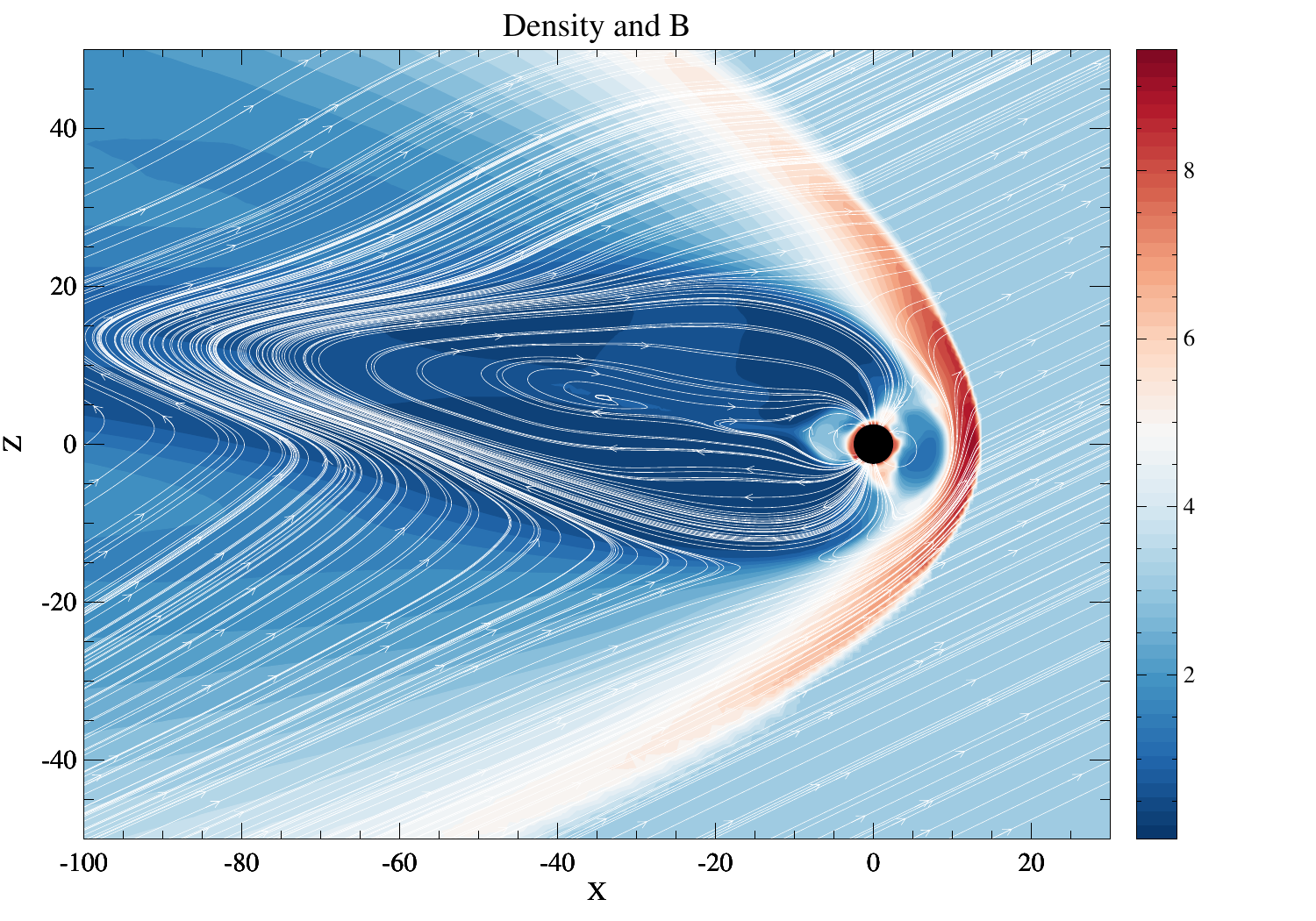}
    \caption{Quasi steady state solution in the meridional cut plane for Earth's magnetosphere after 2,500 steps with local time stepping. Streamlines of the magnetic field are plotted on top of density contours. It is interesting 
    that a reconnection site is present in the stretched magnetotail even for northward pointing interplanetary magnetic field.
    }
    \label{fig:earth_steady}
\end{figure}

\begin{figure}
    \centering
    \includegraphics[width=\linewidth]{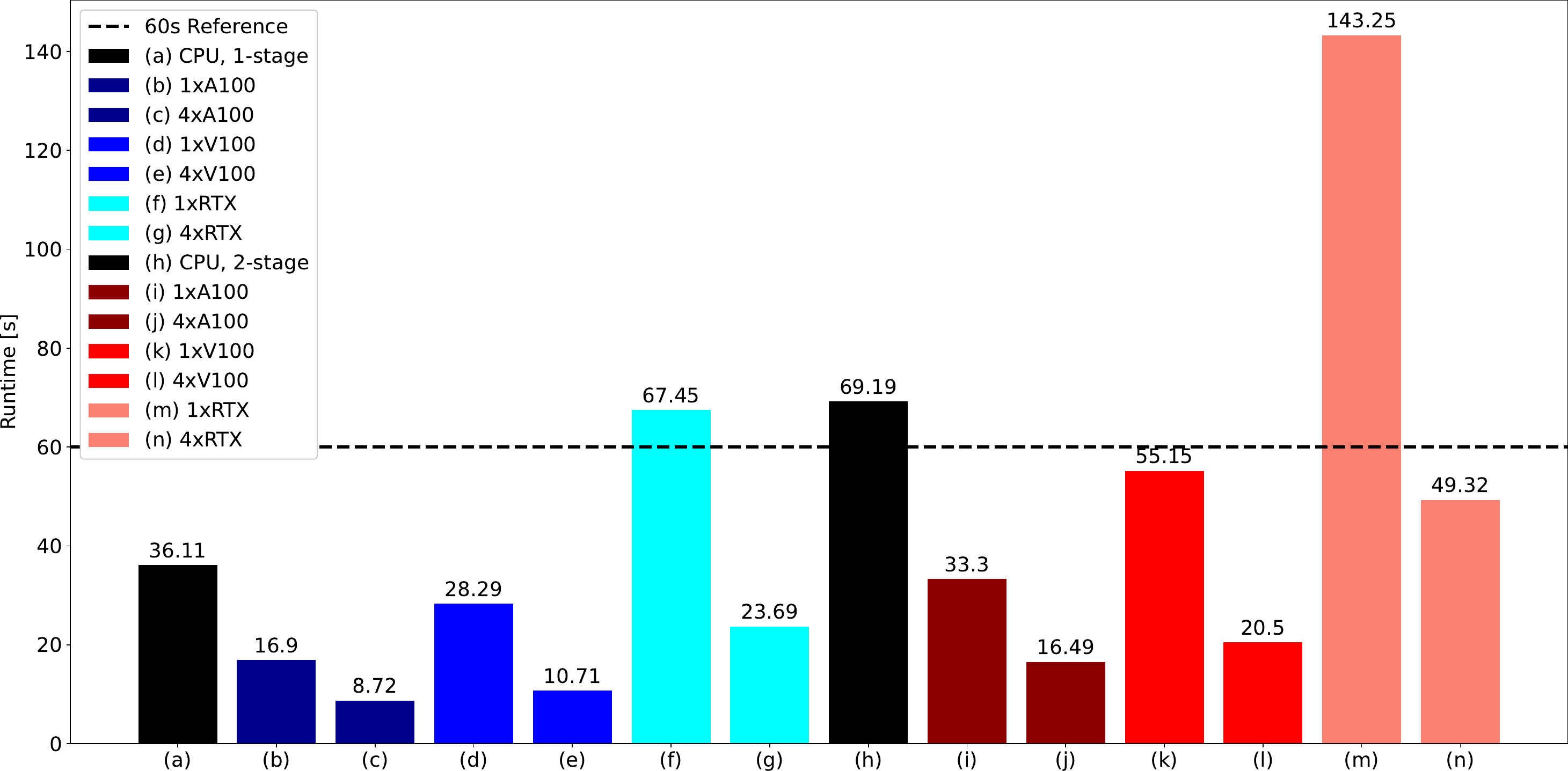}
    \caption{Runtime for 60 seconds of simulated time with various hardware configurations and time stepping settings. Black: CPU reference timings; Blue: 1st order 1-stage time stepping; Red: 2nd order 2-stage time stepping.}
    \label{fig:earth_timing}
\end{figure}

With Earth lying at the origin, the computational domain is the space $[-224,32]\times[-128,128]\times[-128,128]$ in units of Earth radii, $R_E$. Initially, the domain undergoes 8 levels of refinement, which provide better resolution in the close tail region and near the bow shocks. In general, the resolution improves the closer it is to Earth. The resulting grid consists of 3,788 grid blocks ($n_I=n_J=n_K=8$) and around 2 million grid cells in total. The boundary in the $+x$ direction is prescribed with solar wind data obtained on May 4th, 1998, while the others are floating (zero gradient). 
The inner boundary is Earth's ionosphere, modeled by a sphere with radius $3R_E$. To obtain a steady state magnetosphere solution, BATSRUS advances 2,500 steps with local time stepping by the semi-relativistic MHD equations \citep{Gombosi:2002}.
In the time-accurate session that follows, the $+x$ boundary uses a time series of satellite data, and the simulation runs until the simulated time reaches 1 minute.
The steady state session uses second-order time stepping with a CFL number of 0.6. In various runs, the time-accurate session either switches to first-order 1-stage time stepping with a CFL number of 0.85, or uses a temporally second-order two-stage scheme also with CFL=0.85.
Both sessions use a Boris correction factor of 0.02 \citep{Gombosi:2002}. 
The size of the grid and other settings in the 1-stage runs are the same as those used in the Geospace model running operationally at the Space Weather Prediction Center (SWPC), so the timing results are meaningful for production-level simulations.
The 2-stage runs are intended to demonstrate the potential to run higher-fidelity and more accurate simulations faster than real time as well.
The overall workload of the 2-stage scheme is around twice that of a 1-stage time stepping.
Figure~\ref{fig:earth_steady} shows the steady state magnetosphere solution in the meridional plane, and 
Figure~\ref{fig:earth_timing} shows the timings of runs on various hardware.



In addition to the A100 and V100 GPUs on Pleiades, we run the simulation on one 128-core AMD EPYC 7742 ``Rome'' processor to obtain reference CPU timings (using the refactored code and the ifort compiler), and also on the GPU nodes on the Frontera supercomputer.
Each GPU node on Frontera has 4 NVIDIA Quadro RTX 5000 GPUs.
Either 1-stage or 2-stage simulations can run faster than real time on all types of hardware.
Notably, 1-stage simulations can run 3.6 times faster than real time on one A100 GPU and 6.9 times faster on one A100 node.
2-stage simulations can still finish 1.8 times faster than real time on a single A100 GPU and can run 3.6 times faster on one A100 node.
For less powerful GPUs such as the RTX 5000, we are able to run the simulations faster than real time by increasing the number of GPUs.
Compared with the reference CPU timings, one A100 GPU is faster than 270 ``Rome'' CPU cores (2.1 nodes), while one A100 GPU node is faster than 530 ``Rome'' CPU cores (4.1 nodes). 
One V100 GPU is faster than 160 cores (1.3 nodes), and one V100 node is faster than 430 cores (3.4 nodes).
The speedup is similar for both time-stepping settings.

The possibility of running faster-than-real-time magnetospheric simulations on GPUs is meaningful. 
As it only requires a few GPUs to forecast space weather with a practical window, access to supercomputers is no longer a must.
Individuals and institutions can set up a small-scale and affordable machine for such tasks.
It also lays a solid foundation for porting the whole Geospace model \citep{Pulkkinen:2013} to GPUs. A Geospace simulation that couples the aforementioned Global Magnetosphere (GM) model to the Ionospheric Electrodynamics (IE) and the Inner Magnetosphere (IM) models can run about 2.2 times faster than real time on 4 V100 GPUs and 5 CPU cores. We expect further improvements by further optimizing the algorithm.

Another work in progress that benefits from the large speedup of multi-GPU runs is porting the Alfv\'{e}n Wave Solar atmosphere Model (AWSoM)\citep{vanderHolst:2010,vanderHolst:2014awsom} to GPUs. AWSoM simulations are typically run on 2048 CPU cores. 
Based on the parallel efficiency shown in Figure~\ref{fig:scaling}, we expect to be able to run at the same speed on 4-5 A100 nodes.

\section{Conclusions}
\label{sec:Conclusions}
In this work, we have successfully ported a major part of BATSRUS to one and then multiple GPUs. We have modified the original code to minimize data movement between the CPU host and the GPU, changed loop ordering to keep the workload on the GPU adequate, and rewritten the most used parts into a new solver optimal for running on GPUs. The resulting single-GPU code is as fast as the original code running on 270 (A100) or 160 (V100) ``Rome'' CPU cores. 

We have described major modifications to the original message passing algorithm, which make the code scale to multiple GPUs. 
The GPU parallel code has good weak scalability up to 256 A100 GPUs for small workloads, at which point it retains 50\% to 60\% parallel efficiency when compared with a highly optimized single GPU code. 
For large workloads, the code scales well within one computation node, with a parallel efficiency as high as 95\%.
Parallel efficiency drops for large problems run on more than one node because of hardware limitations.
Most importantly, the GPU parallel code can run production-level simulations much faster than real time. 
A single A100 GPU is 3.6 times faster, and a node of four A100 GPUs is 6.9 times faster, which could only be achieved with a much larger number of CPU cores.

This work reveals a hardware bottleneck in large-scale numerical simulations on GPUs: the intra-node, GPU-to-GPU communication bandwidth is an order of magnitude higher than the GPU-to-CPU and inter-node bandwidth. 
This leads to the disparity in efficiency for running large problems on various numbers of GPUs. 
We hope newer supercomputers will have better connectivity between GPUs on different nodes.
Equally importantly, this work enables a more comprehensive space weather forecast (the Geospace model), and prepares us for running other advanced models (AWSoM) efficiently on GPUs.




\section*{Acknowledgment}

The authors acknowledge support from the National Science Foundation grant PHY-2027555. 
We thank the Nvidia Fortran compiler development team, in particular Brent Leback, for their help with resolving issues and improving the efficiency of the OpenACC code.
Computational resources were provided by NSF on the Frontera supercomputer at the Texas Advanced Computing Center and by NASA on the Pleiades supercomputer. BATSRUS is available on GitHub (\url{https://github.com/SWMFsoftware/BATSRUS}), along with instructions for installing the software and generating the manual. A dataset containing the input files and raw timing results can be found on Deep Blue Data (\url{https://deepblue.lib.umich.edu/data/concern/data_sets/ng451j46t?locale=en}).
We acknowledge the use of AI-powered editing service provided by Writeful (\url{https://www.writefull.com}) in revising this manuscript.



\bibliography{csem,mybib}{}
\bibliographystyle{aasjournal}

\end{document}